\documentclass[apj]{emulateapj}

\newcommand{\SII}{\mbox{[S II]}}

\newcommand{\rion}[2]{{\ensuremath{\mbox{\rm #1$\,${\small\uppercase\expandafter{\romannumeral#2\relax}}}}}}

\shorttitle{Herbig-Haro Shocks}
\shortauthors{Dopita \& Sutherlnd}

\begin{document}

\title{Effects of Pre-ionisation in Radiative Shocks II: \newline Application to the Herbig-Haro Objects}

\author{Michael A. Dopita \altaffilmark{1}, Ralph S. Sutherland \altaffilmark{1}} 

\altaffiltext{1}{Research School of Astronomy and Astrophysics, Australian National University, Cotter Road, Weston Creek, ACT 2611, Australia}

\begin{abstract}
In an earlier paper we treated the pre-ionisation problem in shocks over the velocity range $20 < v_{\rm s} < 1000$\,km/s in a fully self-consistent manner. Here we investigate in detail the effect of the upstream UV photon field generated in the radiative zone of shocks in the range in which hydrogen is only partly ionised ($20 < v_{\rm s} < 150$\,km/s). We show that, as a result of super-heating in the non-equilibrium pre-shock plasma, both the magnetic parameter and the Mach number of the shock is strongly affected by the pre-ionisation state of the gas which controls to a large extent the radiative spectrum of the shock. We use these models to provide specific line diagnostics for Herbig-Haro objects which allow us to solve for both the pre-shock density and shock velocity, and present detailed models of the  HH34 jet which allows us to derive the shock conditions, mass-loss rate, momentum flux and chemical abundances in the jet. We show that the refractory elements, Mg, Ca, Fe and Ni are enhanced by 0.22 dex over the solar values, which provides interesting clues about the jet launching mechanism in  pre-main sequence evolution.
\end{abstract}


\section{Introduction}\label{sec:intro}
The structure and predicted emission line spectrum of radiative atomic J-shocks is critically dependent on the ionisation state and the magnetic pressure of the gas entering the shock. For fast shocks \citep{Dopita96,Allen08}, the ionisation of this gas is entirely determined by the EUV photons produced in the cooling zone of the shock, which run ahead to fully pre-ionise the hydrogen in the incoming medium. For shocks with velocities $v_s \gtrsim 150$\,km/s, we can approximate the ionisation state and temperature of the gas entering the shock by an equilibrium photoionisation computation using the upstream EUV photon field.

This approximation breaks down at lower velocities when the upstream photon flux is insufficient to provide full pre-ionisation. In this regime, a useful parameter to describe the pre-ionisation is the  \emph{shock--precursor} parameter, $\Psi$ , introduced by \citet{Shull79}. This is related to the dimensionless ionisation parameter ${\cal U}$ commonly used in photoionisation modelling, which is the ratio of the number density of ionising photons to the atom density.  This can be also written as ${\cal U} = {\cal Q}/c$, where ${\cal Q}$ is the ionisation parameter in terms of the ratio of ionising photons cm$^{-2}$s$^{-1}$ to the particle density (cm$^3$).  To first order  ${\cal Q}$ can be thought of as the velocity of an ionisation front that would be driven into a neutral medium by an ionising radiation field. We can compare this ionisation parameter in the upstream radiation field generated by the shock to the  shock velocity $v_s$ to form the dimensionless shock--precursor parameter: 
\begin{equation}
\Psi = {\cal Q}/v_s = {\cal U}(c/v_s) \, .
\end{equation}

When $\Psi < 1$, the shock velocity is greater than the velocity of the precursor ionisation front.  The photons are trapped by the flux of incoming neutral particles, and there is only a finite time for ionisation and recombination between the moment when the particles enter the precursor region and when they enter the shock. Since the precursor partly-ionised zone in slow shocks is relatively thin, the timescale for incoming atoms to be advected through this zone can be considered short compared to the recombination timescale. In this case, to a good approximation, we can simply balance the upstream EUV photon flux with the number of ionised hydrogen atoms advected into the shock, $xn_{\rm H}v_s$, where $x$, is the fractional ionisation of hydrogen in the gas entering the shock. Thus, to a good approximation, $\Psi \sim x$. 

In the first paper of this series (Sutherland \& Dopita, 2017; hereinafter Paper I), we provided a fully self-consistent modelling of both the pre-shock ionisation and heating over the full velocity range covered by non-relativistic radiative atomic J-shocks, and we investigated how the pre-shock ionisation is influenced by the transverse magnetic field. For the slower shocks which are unable to establish an equilibrium photoionisation region ahead of the shock, we identified three distinct shock regimes:
\begin{itemize}
\item{{ \it Cold Neutral Precursors} ($v_s \lesssim 40$\,km/s). Here the upstream precursor radiation field is both too weak and too soft, to either heat or ionise the gas entering the shock front.}
\item{ {\it Warm Neutral Precursors} ($40 \lesssim v_s \lesssim 80$\,km/s). Here the precursor radiation field contains enough high energy photons, at a very low ionisation parameter, to produce energetic electrons which can heat the gas significantly.  While the total ionisation remains fairly low, the heating per ionisation rises, and this heating is not balanced by efficient electron collisional cooling, so the electron temperature can rise to well over $10^4$ K.}
\item{ {\it Warm Partly-Ionised Precursors} ($80 \lesssim v_s \lesssim 140$\,km/s). Here $\Psi < 1.0$, so hydrogen is partially pre-ionised. However, hydrogen remains less ionised when entering the shock than would be the case if the gas had time to come into equilibrium with the precursor field. }
\end{itemize}

\begin{figure}
\centering
  \includegraphics[scale=0.6]{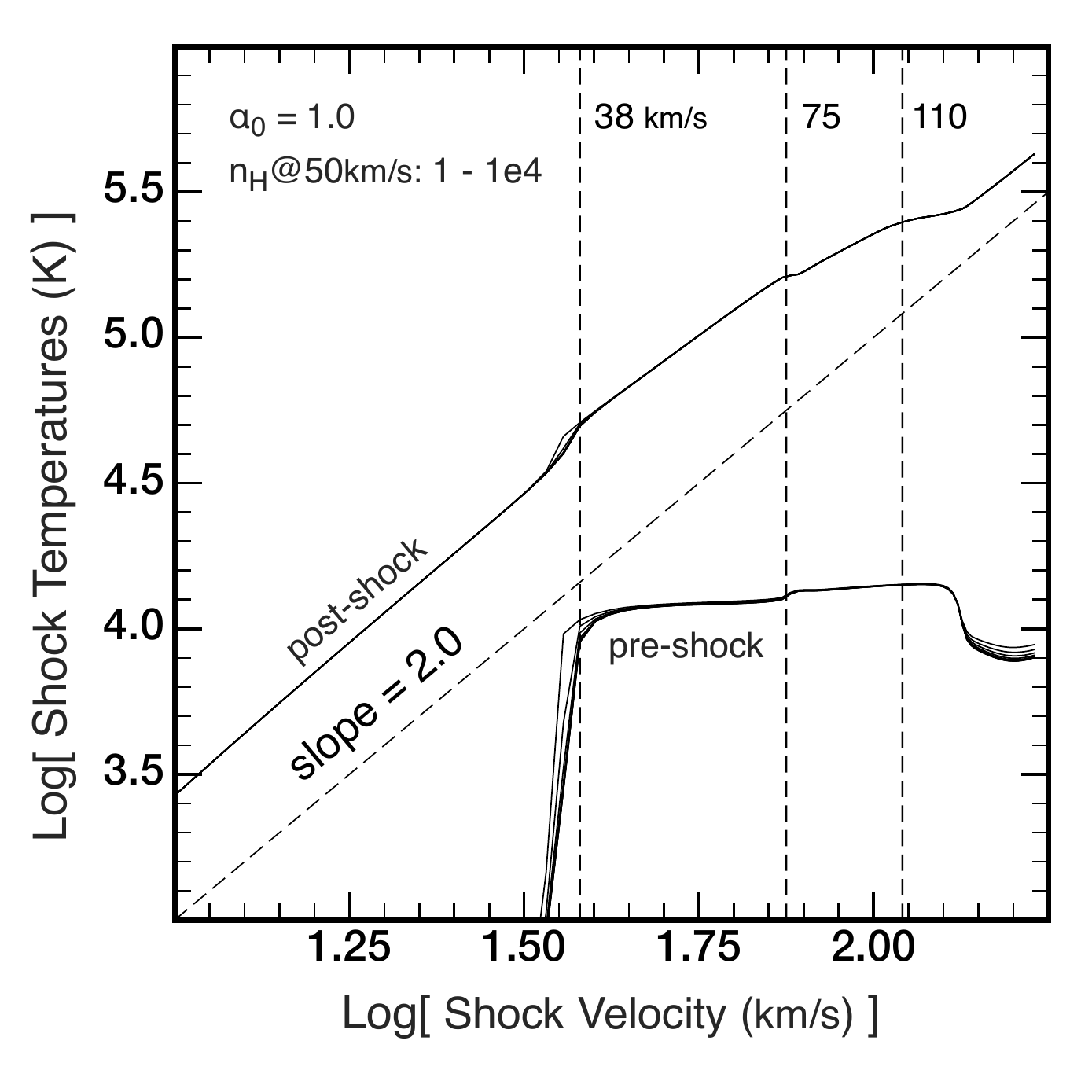}
  \caption{The pre- and post-shock temperatures as a function of shock velocity. The different curves correspond to different pre-shock densities ($1 \leqslant n_{\mathrm H} \leqslant 10000$cm$^{-3}$). These are virtually degenerate. The deviations of the post-shock temperature with velocity from the slope = 2.0 (shown as a dashed line) are fundamentally driven by changes in the molecular weight of the pre-shock plasma. The exception is at around $v_s \sim 38$km/s, where the sudden increase in the pre-shock temperature changes the Mach number of the shock.} \label{fig:shock}
 \end{figure}

These three regimes can be readily identified if we consider the pre- and post-shock temperatures as a function of velocity, as shown in Figure \ref{fig:shock}. Note that the post-shock temperature overall has a slope of +2, but deviates from this due mostly to changes in the pre-shock molecular weight. Note also that the pre-shock temperature in the warm, partly-ionised shock precursor regime is higher than in fast shock regime( $v_s > 120$\,km/s) as a result of superheating of the gas in the precursor.

In the Herbig-Haro (HH) objects, which are shocks generated by outflows in pre-main sequence stars such as T Tauri and FU Orionis stars, we find examples covering all of these three regimes. Thus, for the HH objects, the issue of pre-ionisation is critical in the interpretation of their spectra. In this paper, we will examine in more detail the role of shock pre-ionisation in controlling the emission line spectrum of HH objects, and will endeavour to define diagnostic tools which can aid both in the interpretation of spectra, and the derivation of the physical parameters.

\section{The Herbig-Haro (HH) Objects}
A quantitative interpretation of the spectroscopy of the Herbig-Haro (HH) objects is capable of providing great insights into the formation, mass accretion and mass ejection by pre-Main Sequence stars. Many of their spectroscopic and dynamical properties have been summarised in a series of excellent reviews \citep{Schwartz83, Edwards93, Reipurth95, Reipurth97, Bohm97, Reipurth01, Bally02a}

Since \citet{Schwartz75} recognised that the HH objects are shock excited, there have been many attempts to explain their spectra using both radiative shock modelling \citep{Dopita78, Bohm80, Dopita82a, Dopita87, Raga95, Raga96} and hydrodynamic modelling \citep{Hartmann84, Raga85, Hartigan87, Biro94, Henney96}.

As first pointed out by \citet{Dopita82b}, the HH shocks arise in a dense, two-sided jet thrown out by an eruptive low-mass pre-main sequence star. These jets may extend over parsec scales \citep{Heathcote98, Bally02a, Bally02b} and can interact with the surrounding ISM in a variety of ways. In some objects dense ``bullets'' of gas are episodically ejected, producing fast bow-shocks and slower cloud shocks. In others, the shocks are produced as weak internal jet shocks, which generally have low internal shock velocities and correspondingly low excitation, but which display very rapid transverse velocities \citep{Bally02b,Hartigan05, Beck07, Raga16}. 

The HH objects display a wide range of excitation ranging from the extremely low-excitation objects described by \citet{Bohm80}, through the rich spectrum displayed by objects such as HH1 \citep{Solf88}, up to extremely high excitation objects such as HH80 and 81 \citep{Heathcote98}, which have spectra which are indistinguishable from the fast radiative shocks found in supernova remnants.  Generally, the \SII\ ratios indicate that the pre-shock densities are high, and the strength of the [Fe\,II] lines as well as of the Ca\,II] and Mg\,I] lines \citep{Raga96, Reipurth00b} indicates that dust has largely been destroyed in the shock \citep{Beck-Winchatz96}. The (then) available optical spectrophotometry on the HH objects was reviewed by \citet{Raga96}.

The low-excitation HH objects present a particularly interesting optical spectrum. In these, lines such as [C\,I] $\lambda\lambda 9823,9849$ can exceed the strength of H$\alpha$, as can [O\,I] $\lambda6300$, while [N\,I] $\lambda\lambda 5198,5200$ can exceed H$\beta$ in intensity. In addition, these objects display a large H$\alpha$/H$\beta$ ratio, the absolute surface brightness of H$\beta$ is low, and the 2-photon continuum of hydrogen can become relatively very strong \citep{Bohm80, Dopita82a}. Such properties can only be explained in terms of low-velocity shocks with a low pre-ionisation fraction of hydrogen. In the following sections, we will examine how our new self-consistent models can inform us about the physical conditions prevailing in such objects.

\section{Shock Diagnostics for HH Objects}
It should be noted at the outset that shock models adopting a single shock velocity and pre-shock density will fail to reproduce the observed spectra of HH objects. This is because ground-based spectrophotometry includes a wide range of shock conditions within the aperture used for the observation. We may recognise the presence of a wide number of shock types. These include bow shocks \citep{Raga85, Hartigan87, Henney96} and cocoon shocks \citep{Heathcote98} propagating in the surrounding ISM, in addition we may find cloud shocks, jet termination shocks and working surface shocks internal to the jet \citep{Raga95,Reipurth97,Bally02a, Beck07}. To separate these requires high spatial resolution integral field spectrophotometry of which, apart from the studies by \citet{Beck07} and \citet{Lopez10}, there has been very little. 

\subsection{Model Grid}
To provide a set of shock models which can be used to derive shock diagnostics for the physical conditions prevailing in HH objects, we have run a dedicated series of models using Local Galactic Concordance (LGC) abundances (Nicholls et al. 2016, MNRAS, in press), and covering the velocity range $15 \leqslant V_s \leqslant $\,150km/s, in steps of 2.5\,km/s at fixed shock ram pressure. This eliminates the primary velocity-dependence of the surface brightness, and is more applicable to real objects in which the pressure is the key parameter which drives shocks of varying velocity through a multiphase medium, or in the case of a cloud driving into the ISM, which produces a fast bow shock and a slower cloud shock with common ram pressure.

Given that \citet{Beck-Winchatz96} finds that, in most HH objects, the Fe abundances seem to be un-depleted in the gas phase, as our primary HH grid we adopted dust-free models. We have also run a depleted set with [Fe/H] = -1.0, where the appropriate depletion factors for the individual refractory elements are taken into account, as described in \citet{Dopita16}. However, these models do not provide a good fit to the observed HH object spectra. The ram pressure, $P=\rho V_s^{2}$ is normalised to the pre-shock hydrogen number density at a shock velocity of 50\,km/s. For a pre-shock hydrogen density  $n_1 = 1.0$\,cm$^{-3}$, this corresponds to a pressure $\log P = -10.27$\,dynes cm$^{-2}$. Constant pressure shock families in the range $15 \leqslant V_s \leqslant $ 150km/s were run for the pre-shock hydrogen densities of $n_{\mathrm H} = 1, 3, 10, 30, 300, 1000, 3000$ and 10000 cm$^{-3}$. 

As for the magnetic field, we have assumed that equipartition between the magnetic and the gas pressures pertains in the proto-ionised state, and that both the particle density and the magnetic field remain constant in the pre-ionised region. Thus, in these models, the magnetic field no longer dominates the gas pressure when the pre-ionised gas has been preheated. The main effect of the magnetic field is to limit the compression in the cooling zone of the shock (see Paper I), which affects [O\,II] and [S\,II], but has little effect on other line ratios in our computed velocity range. This effect was clearly shown by \citet{Hartigan94}.

\subsection{Shock Surface Brightness}
\begin{figure*}
\centering
  \includegraphics[scale=0.6]{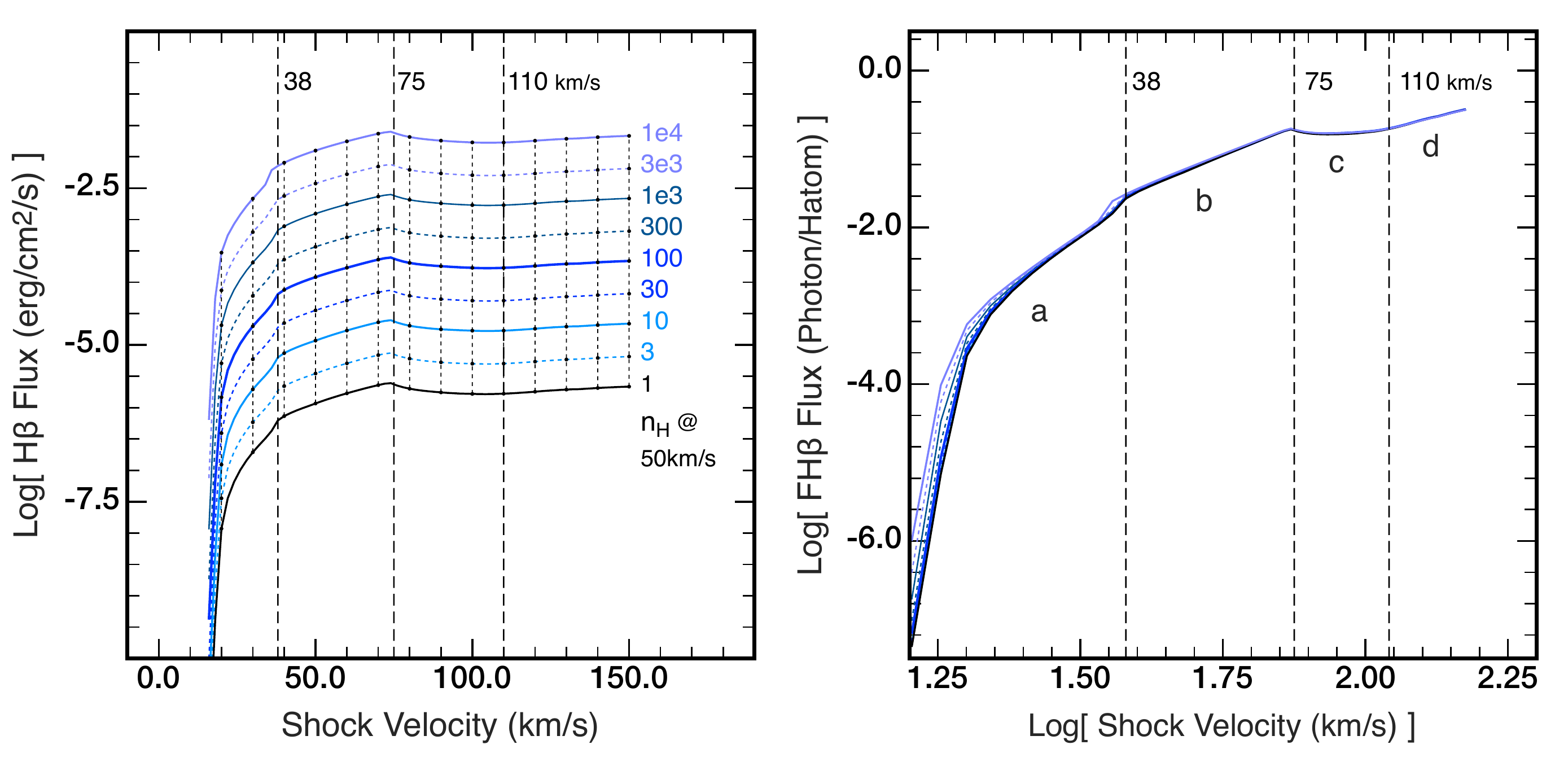}
  \caption{(Left) the surface brightness of the radiative shocks as a function of shock velocity and pre-shock hydrogen number density $n_{\mathrm H}$ in units of  cm$^{-3}$. (Right) As was shown in Paper I, this can be simply scaled as $\Psi_{H\beta}$,  the number of H$\beta$ photons produced per hydrogen atom entering (or advected into) the shock, which almost entirely removes the density dependence. The velocity thresholds between the shock types summarised in the introduction are marked as dashed vertical lines, and fits to the segments a--d are given in Paper I.} \label{fig:SB}
 \end{figure*}

The key factor in determining the observability of a shock is clearly its surface brightness. In Figure \ref{fig:SB} we show the H$\beta$ surface brightness as a function of both ram pressure and shock velocity. Naively, we might expect the H$\beta$ surface brightness to scale as the rate of mechanical energy flux across the shock $\dot E =0.5 \rho v_s^3$. This is clearly not the case. In practice, it is the product of the number of recombinations occurring per hydrogen atom entering the shock  and the rate at which these atoms are processed through the shock which governs the surface brightness. This is a more complex function, but which scales more closely as the ram pressure, $ \rho v_s^2$. For slow shocks, the surface brightness falls rapidly due to the rapidly decreasing mean fractional ionisation of hydrogen within the shock. As a consequence shocks with velocities $v_s\lesssim $30\,km/s are unlikely to be observed. This concurs with the observations of \citet{Bohm80}, who found that the lowest excitation HH objects do indeed have the lowest surface brightness.  Figure  \ref{fig:SB}  shows an increase in surface brightness occurs once the pre-shock gas becomes becomes warm at $v_s \gtrsim 38$\,km/s, followed by a decrease once the pre-ionisation fraction becomes appreciable at $v_s \gtrsim 75$\,km/s, due to the changing mean atomic weight, $\mu$. At the highest velocities, the surface brightness starts to approach our expectation.

 \subsection{Precursor Surface Brightness} \label{sec:SBprecursor}
 While the degree of pre-ionisation remains low, for shock velocities  $v_s < 75$\,km/s, the precursor has a very low surface brightness compared with the shock and remains essentially unobservable. To be seen in the low velocity regime, the shock would have to be externally irradiated by a UV source. Such shocks were first identified by \citet{Bally01}, and have since been extensively studied  both observationally \citep{Bally02c, Bally09, Comeron13, Riaz15} and theoretically \citep{Kajdic07}. However, the models presented here are not appropriate for such objects.
 
 At higher velocities, the precursor emission becomes more significant relative to the optical emission in the main shock, and finally, when the shock is fast enough to supply soft X-rays to the precursor region ($v_s \gtrsim 250$\,km/s) the precursor emission is sufficient to profoundly influence the global shock + precursor spectrum \citep{Dopita95}.
 
\begin{figure*}
\centering
  \includegraphics[scale=0.6]{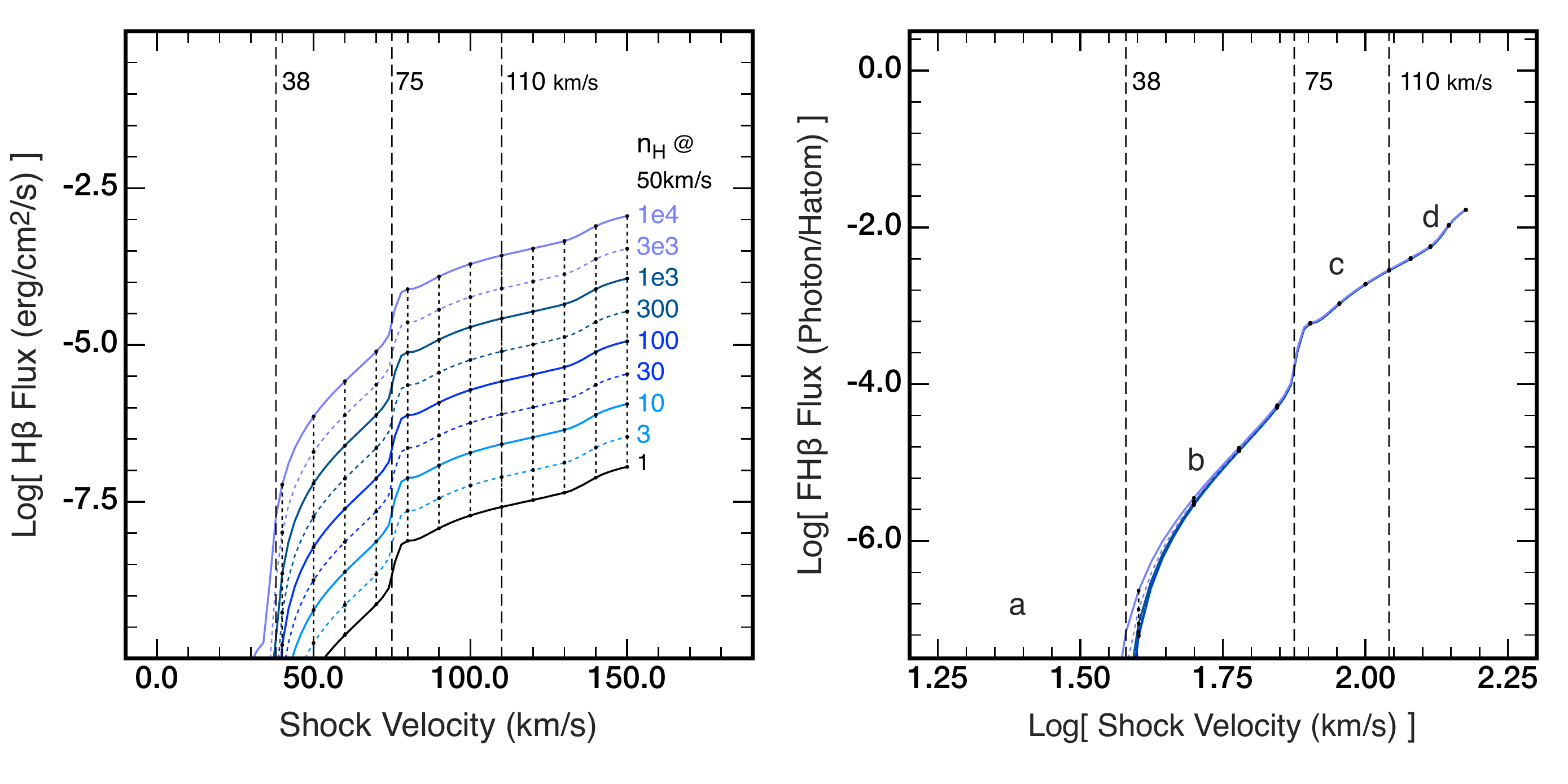}
  \caption{As Figure \ref{fig:SB} but for the shock precursor. Left: the surface brightness of the precursors as a function of shock velocity and pre-shock hydrogen number density $n_{\mathrm H}$ in units of  cm$^{-3}$.  Right: the number of H$\beta$ photons produced per hydrogen atom entering (or advected into) the shock. The velocity thresholds between the shock types summarised in the introduction are marked as dashed vertical lines.} \label{fig:SBprec}
 \end{figure*}

 In Figure \ref{fig:SBprec} we show the surface brightness of the precursor in the range of velocities relevant to the HH objects. This should be compare with the surface brightness of the shocks given above in Figure \ref{fig:SB}. In the regime where the precursor is cold, and essentially un-ionised, the contribution of the precursor can be entirely neglected. Even at the upper limit of the HH grid ( $v_s = 150$\,km/s), the brightness of the precursor only amounts to a few percent of the brightness of the radiative shock, and so remains rather difficult to observe.
 
\begin{figure}
\centering
  \includegraphics[scale=0.6]{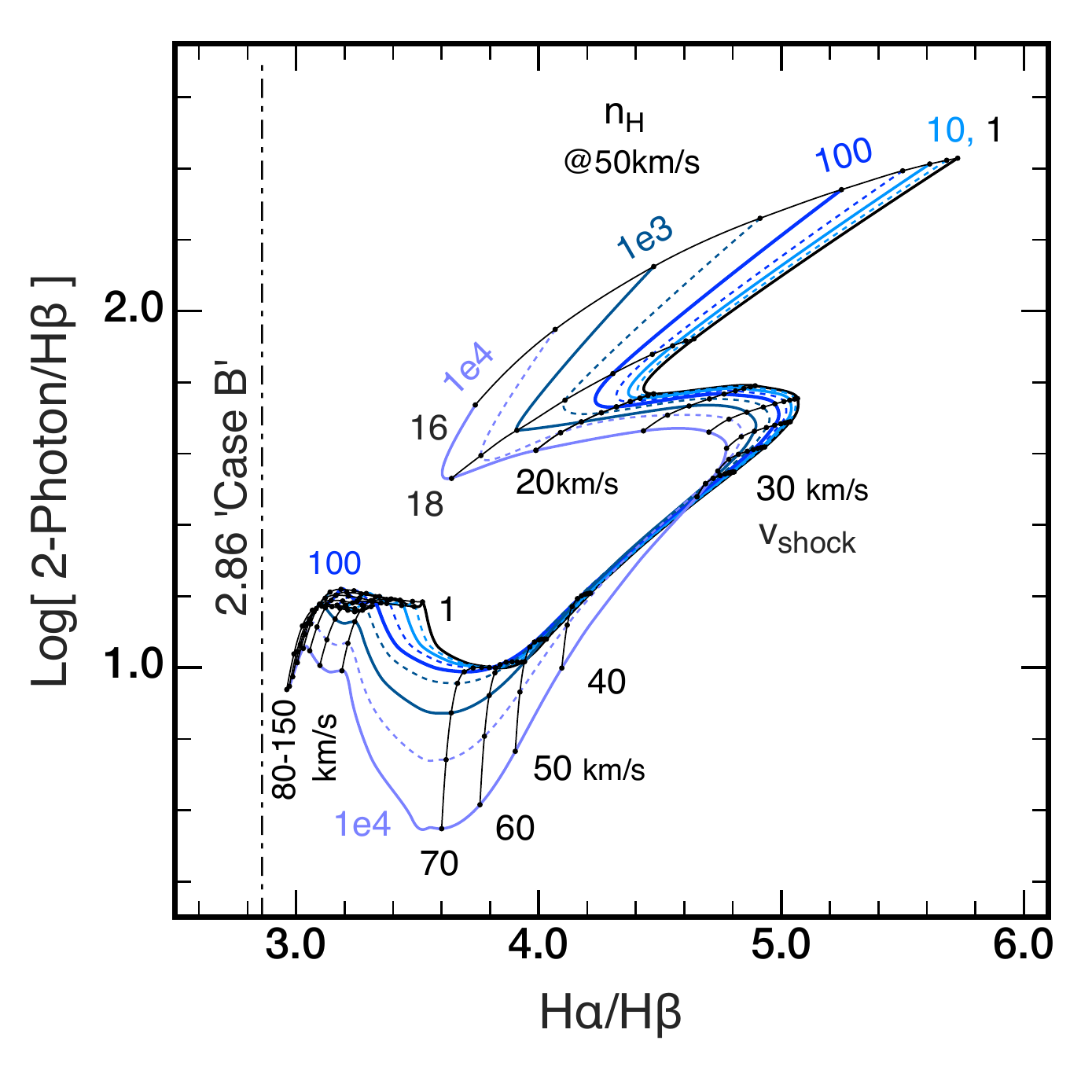}
  \caption{(From Paper I) The relationship between the Balmer decrement, H$\alpha$/H$\beta$, and the hydrogen 2-photon continuum. Note that it is possible to obtain an enhancement of the 2-photon continuum by a factor greater than 30. In all shocks, the intrinsic Balmer Decrement is larger than the photoionised nebula Case B value, 2.86 (marked here as a vertical dashed line).} \label{fig:twoquantum}
 \end{figure}

\subsection{Balmer Decrement and Hydrogen 2q Emission}\label{sec:2q}
\citet{Dopita82a} discovered that many low-excitation HH objects show a pronounced enhancement of the hydrogen two-photon (2q) continuum, accompanied by an increase in the intrinsic H$\alpha$/H$\beta$ ratio. They explained this in part in terms of slow-shocks moving into a medium with low pre-ionisation of hydrogen. However, their models  were limited at F(H$\alpha$)/F(H$\beta) \sim 3.9$ and F(2q)/F(H$\beta) \sim 60$. To obtain the maximal values observed (F(H$\alpha$)/F(H$\beta) \sim 4.6$ and F(2q)/F(H$\beta) \sim 120$), \citep{Dopita82a} needed to additionally invoke finite-age shocks. 

Our current self-consistent steady-flow shock models produce flux ratios which naturally reach these maximal values observed in HH objects without having appeal to finite age shocks, provided that the shock velocities are very low -- of order 15-20\,km/s (see Figure \ref{fig:twoquantum}). Less extreme enhancements of F(2q)/F(H$\beta$) occur in all shocks with $v_s < 45$\,km/s, but all models with less than complete pre-ionisation display enhanced Balmer decrement due to collisional excitation of the Hydrogen $n=3$ level. If the pressure driving the shock does not vary by too much object to object, Figure \ref{fig:SB} also implies that the shock surface brightness should be also correlated with Balmer decrement and the 2q excess, provided that the pressure driving the shock does not vary by too much object to object. This would help explain the correlation noted by \citet{Bohm80} in their Figure 1.

A different way of producing enhanced H$\alpha$/H$\beta$ ratios in HH objects was recently identified in HH1 and 2 by \citet{Raga15b}. Using HST imaging, they found strong enhancements in this ratio immediately behind the bow shock of these objects. In this regard, the HH objects are behaving in the same way as the Balmer-dominated shocks seen in supernova remnants. These require that the gas  entering the shock has not passed through an ionised precursor (which implies either finite or partially-radiative shocks) resulting in strong collisional excitation of hydrogen in the region where it is first becoming ionised. The theory of such `collisionless' shocks was originally developed by  \citet{Chevalier78}.

\subsection{Velocity Diagnostics}
The observations of HH objects summarised by \citet{Raga96} display a paradox that has been encountered in all earlier attempts to derive shock parameters, namely, that the strength of the low-excitation lines [O\,I], [N,I] and [S\,II] implies very low shock velocities, while the simultaneous presence of higher-excitation lines such as [N\,II] and [O\,III] requires higher shock velocities;  above 50\,km/s in the case of [N\,II] and above 80\,km/s for [O\,III]. The root cause of this discrepancy appears to be that the HH shocks are - in general - composite, arising from the summation of a jet shock of some kind (internal working surface, jet density pulse or jet Mach disk) and an external bow or cocoon shock of both higher velocity and lower density. For this reason, shock diagnostics which depend on a limited set of line ratios must be treated as indicative, rather than quantitative. Furthermore, much of the spectrophotometry on the HH objects is old, and of limited accuracy by today's standards, and reddening corrections are heterogeneous, depending on the reddening ``law" adopted. In addition, reddening corrections are often based upon correcting observed Balmer decrements to the standard Case B decrement, while Section \ref{sec:2q}, above, makes it clear that such an assumption is not generally valid. However, within these caveats, let us proceed regardless to determine what physical parameters we can estimate using the existing data from \citet{Raga96} and \citet{Heathcote98}.

\begin{figure*}
\centering
  \includegraphics[scale=0.6]{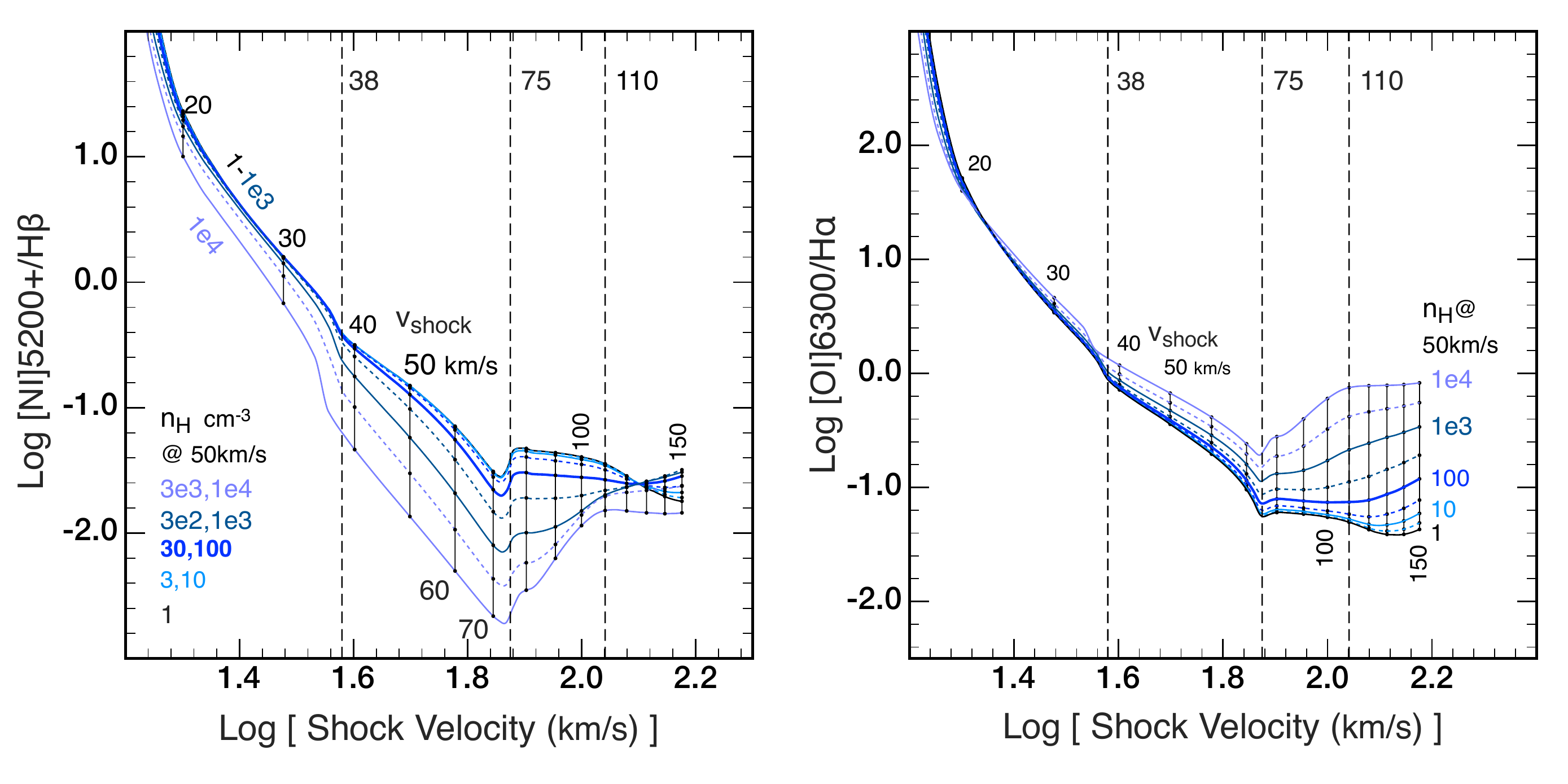}
  \caption{The variation of [(left) the N\,I]$\lambda5200$/H$\beta$ and [(right) O\,I]$\lambda6300$/H$\alpha$ ratios with shock velocity and pre-shock density.} \label{fig:tNIOI_diagnostic}
 \end{figure*}

In principle, the shock velocity could be derived from the observed [N\,I]$\lambda5200$/H$\beta$ or [O\,I]$\lambda6300$/H$\alpha$ alone. This has been previously done in the extensive grid of shock models by \citet{Hartigan94}, who investigated the effect of high pre-shock magnetic fields as well as high density in HH shocks. In Figure \ref{fig:tNIOI_diagnostic}, we show the results for our self-consistently pre-ionised and pre-heated set of models. Our models show very little density dependence for the slower shocks. This is driven by the fact that these have very low fractional pre-ionisation, and low maximum ionisation fractions as well, so all slow shock models remain at, or close to, the low density limit. At higher velocities, the agreement with the  \citet{Hartigan94} models seems to be quite good. Our models show evidence for collisional de-excitation in [N\,I]. By contrast, the [O\,I]$\lambda6300$/H$\alpha$ ratio is enhanced as a consequence of collisional de-excitation in both [N\,I] and [S\,II].

When comparing these two line ratios with the observations of HH objects we find that there is a disagreement between the two, in the sense that the observed [N\,I]$\lambda5200$/H$\beta$ provides a somewhat lower shock velocity than does the  [O\,I]$\lambda6300$/H$\alpha$ ratio. We simply adopt the mean of these two as the estimated shock velocity using these ratios.

\begin{figure}
\centering
  \includegraphics[scale=0.55]{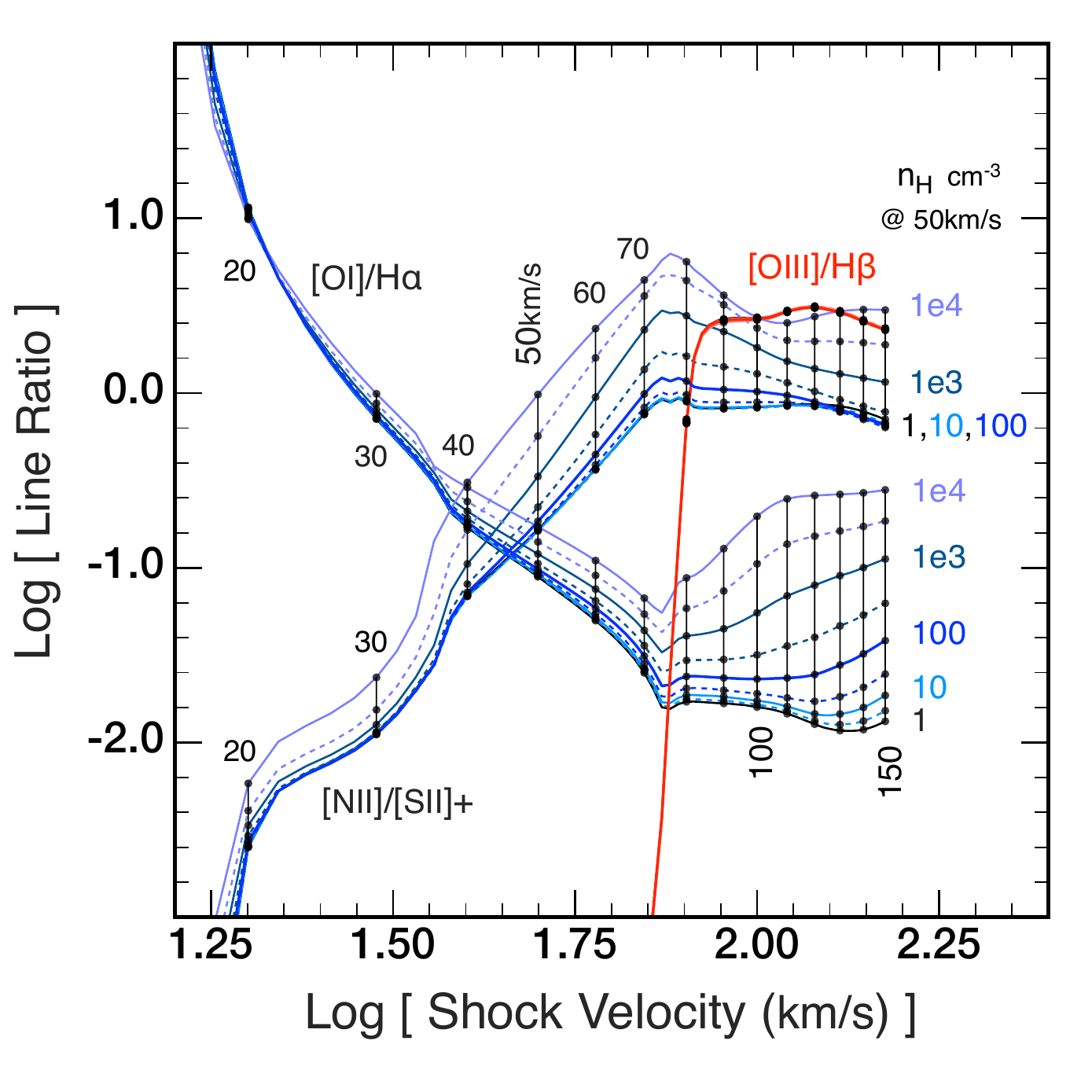}
  \caption{Shock velocity diagnostic ratios derived from the [O\,I]$\lambda6300$/H$\alpha$,  the [N\,II]$\lambda6584$/[S\,II]$\lambda\lambda6717,6731$ and  the [O\,III]$\lambda5007$/H$\beta$ emission line ratios. Note that collisional de-excitation effects and reddening correction effects are partly compensated for by this particular choice of ratios.} \label{fig:vel_diagnostic}
 \end{figure}

Other ratios can also be used to constrain the shock velocity. In Figure \ref{fig:vel_diagnostic} we present what appear to be the most useful of these ratios;  [O\,I]$\lambda6300$/H$\alpha$,  [N\,II]$\lambda6584$/[S\,II]$\lambda\lambda6717,6731$ and [O\,III]$\lambda5007$//H$\beta$. In this particular set, all ratios depend only weakly on the assumed reddening correction. In addition, the  [O\,III]$\lambda5007$//H$\beta$ ratio is effectively independent of density and the density sensitivities of the [O\,I]$\lambda6300$/H$\alpha$ and N\,II]$\lambda6584$/[S\,II]$\lambda\lambda6717,6731$ ratios are similar.

Due to the issues identified above, the observed [N\,II]$\lambda6584$ line strengths are stronger than predicted in a single-velocity shock model. Another factor which may cause this effect is the fact that many HH objects are embedded in weak diffused ionised nebulosity which is characterised by large  [N\,II]$\lambda6584$/H$\alpha$ line ratios. A good agreement between the observed [O\,II]$\lambda6300$/H$\alpha$ and the theoretical [N\,II]$\lambda6584$/[S\,II]$\lambda\lambda6717,6731$ can be obtained if the latter ratio is enhanced by $\sim 0.4$\,dex. With this estimate of the enhancement, we have also used the observational material from \citet{Raga96} and \citet{Heathcote98} to derive shock velocities (ignoring the values given by the [O\,III]$\lambda5007$//H$\beta$ ratio for this exercise).

\begin{figure}
\centering
  \includegraphics[scale=0.55]{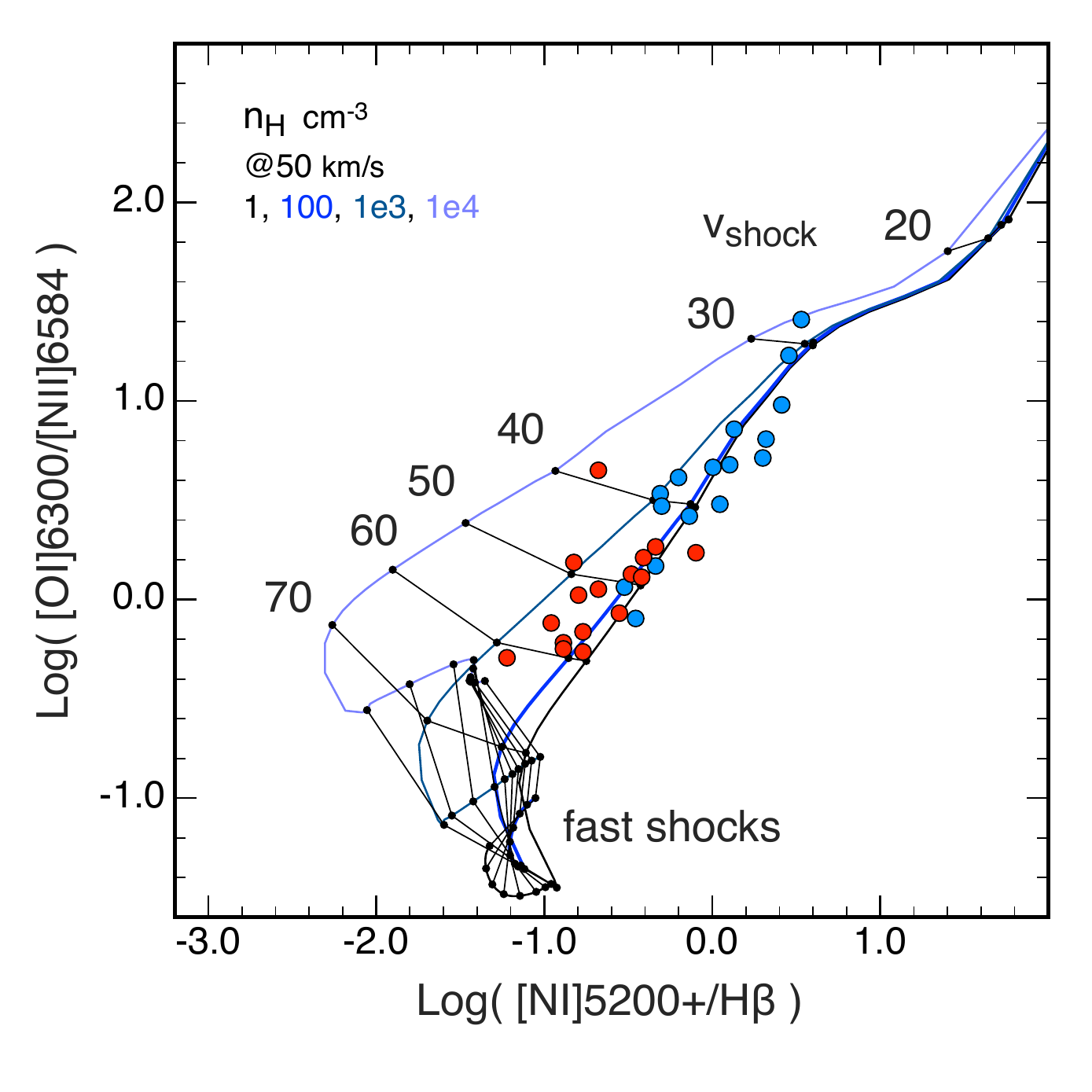}
  \caption{A shock velocity diagnostic derived from the [N\,I]$\lambda5200$/H$\beta$ and the [O\,I]$\lambda6300$/H$\alpha$/[N\,II]$\lambda6584$ ratios. The red points represent those objects with [O\,III]/H$\beta > 0.15$, and the blue points are for objects with slower shock velocities in which the [O\,III] $\lambda 5007$ emission is either weak or absent. } \label{fig:OININII_diagnostic}
 \end{figure}

Finally, in Figure \ref{fig:OININII_diagnostic} we show the [N\,I]$\lambda5200$/H$\beta$ vs. [O\,I]$\lambda6300$/H$\alpha$/[N\,II]$\lambda6584$ ratio, which appears to provide a good velocity discrimination up to a velocity of about 70\,km/s, and against which the observational data can be directly compared. Also shown are the spectrophotometric data from \citet{Raga96} and \citet{Heathcote98} in which we have applied the 0.4\,dex offset to the Nitrogen lines which was estimated above. On the figure, the red points represent those objects with [O\,III]/H$\beta > 0.15$, and the blue points the rest. Clearly, those objects showing the [O\,III] line are of systematically higher velocity than the others, but none are compatible with the $v_s \sim 80$km/s needed to produce this line in the single shock models. Clearly therefore, the spectrophotometric data of the HH objects must reflect the presence of a fairly wide range of shock velocities present within them.

With this offset we have estimated the velocities in the lower-velocity (cloud) shock component of the observed HH objects given in \citet{Raga96} and \citet{Heathcote98}. In all those objects for which [O\,III]$\lambda5007$ is detected,  the observed [O\,III]$\lambda5007$//H$\beta$ ratios imply (bow) shock velocities of close to 75--85\,km/s, while the lower-ionisation lines provide a solution with a lower velocity than this.  The exception to this rule is  HH81, for which a fairly consistent solution with $70 \lesssim v_s \lesssim 85 $\,km/s is obtained using all three line ratios. In Table \ref{Table:HHparms} we list the indicative shock velocities so obtained. We emphasise that these are merely crude estimates, and we have no way of evaluating the error on either the data or the fit. These serve merely to inform future attempts to provide detailed models of individual objects.

\subsection{Estimating the Ram Pressure}
In our models, the density-sensitive [S\,II] ratio $\lambda\lambda 6731/6717$ serves to provide an estimate of the ram pressure in the shock. However, this ratio depends not only on the fractional ionisation of the gas entering the shock, but also on the degree of ionisation reached within the shock, and on the compression factor of the gas passing through the cooling region of the shock. To fit the observations to these self-consistent pre-ionisation and pre-heating shock models we need have a means of disentangling the shock velocity from the ram pressure.

One way of doing this  is to plot  a density sensitive  ratio such as [S\,II] $\lambda\lambda 6731/6717$ against a velocity sensitive ratio. Suitable candidate ratios are [N\,II]$\lambda6584$/[S\,II]$\lambda\lambda6717,6731$  or the [N\,II]$\lambda6584$/[N\,I]$\lambda\lambda5198,5200$. These are shown in Figure \ref{fig:diagnostics}, along with the observational data from \citet{Raga96} and \citet{Heathcote98}. Here, again,  we distinguish the objects with {[O\,III]/H$\beta < 0.15$  (in blue) from those with larger  [O\,III]$\lambda5007$//H$\beta$ emission line ratios (in red). It is clear from the plot that the objects with larger [O\,III]/H$\beta$ ratios are indeed characterised by larger shock velocities. Again, some of these points would indicate shock velocities as low as 35\,km/s, while shock models do not produce appreciable [O\,III] emission until the velocity is in excess of 75\,km/s. Nonetheless, the agreement of the observed points between panels (a) and (b) of Figure \ref{fig:diagnostics} is rather good.

A better way to estimate the ram pressure is to directly use the shock velocity derived above to solve for the ram pressure as indicated by the [S\,II] ratio $\lambda\lambda 6731/6717$ ratio. This is shown in Figure \ref{fig:press_diagnostic}. At low shock velocities, the fractional ionisation of hydrogen in the shock remains low, and high pre-shock densities are required to raise the [S\,II] ratio $\lambda\lambda 6731/6717$ ratio above its low density limit. Typical pre-shock densities for the observed HH objects are in the range $10-1000$\,cm$^{-3}$. The derived velocities, densities and implied ram pressures for individual HH objects are listed in Table \ref{Table:HHparms}.

\begin{figure*}
\centering
  \includegraphics[scale=0.55]{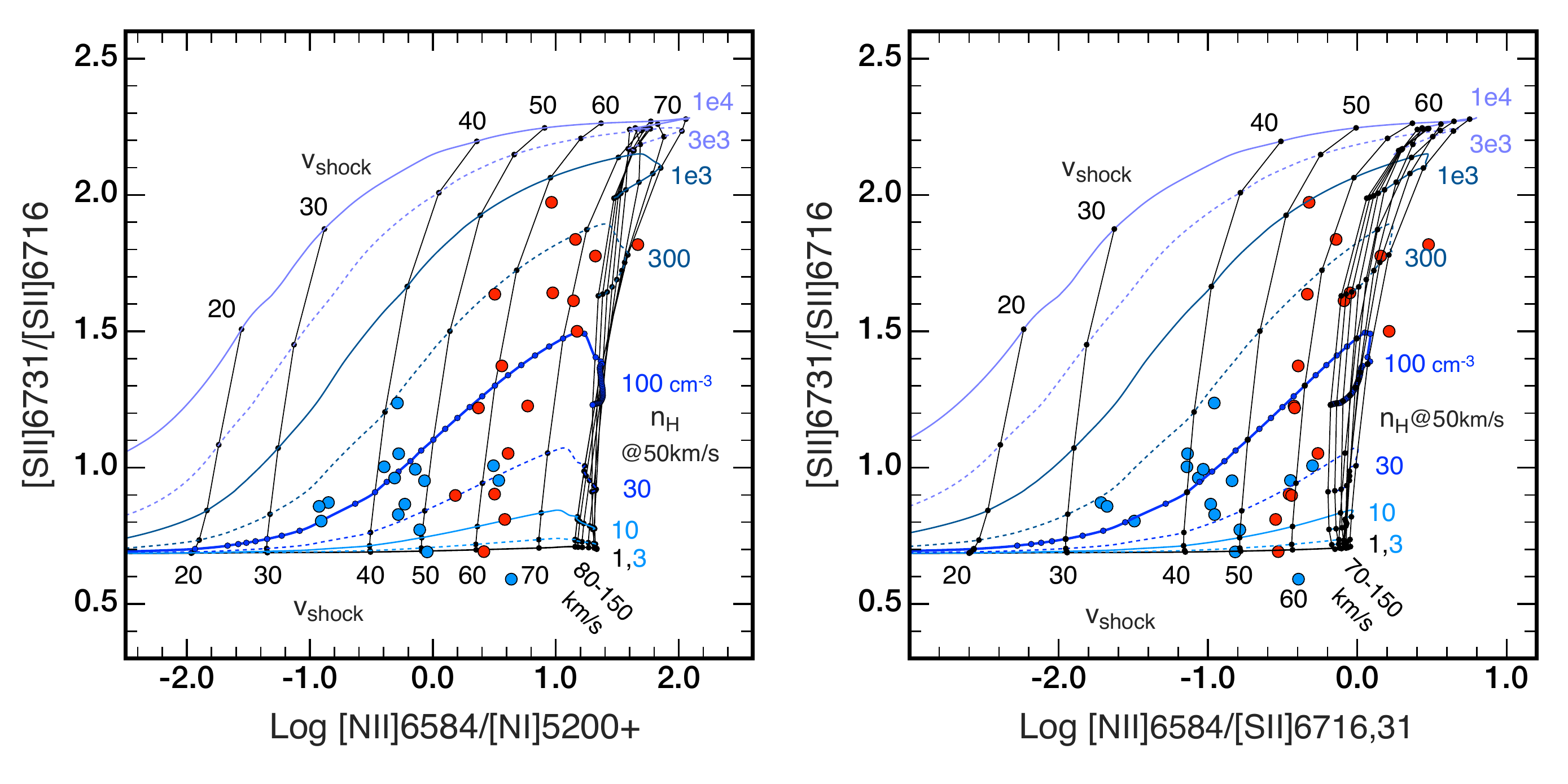}
   \caption{Line diagnostics to designed to provide a simultaneous solution for both shock velocity and shock ram pressure for HH objects. The strength of the [N\,II] line relative to ions of lower ionisation potential is sensitive to the shock velocity below $v_s \sim 80$\,km/s while the [S\,II] $\lambda\lambda 6731/6717$ measures the compression factor in the shock, which is sensitive to the ram pressure (given on the right of each plot in units of dynes cm$^{-2}$). For convenience, we also give the corresponding pre-shock hydrogen density (cm$^{-3}$) for a shock velocity of 50\,km/s. The observed points are derived from the compilation of \citet{Raga96}, supplemented by the subsequent observations of \citet{Heathcote98}. The blue symbols are for those objects displaying [O\,III] $\lambda 5007$ less than 15\% of H$\beta$, and the strong [O\,III] emitters are in red. } \label{fig:diagnostics}
\end{figure*}

\begin{figure}
\centering
  \includegraphics[scale=0.55]{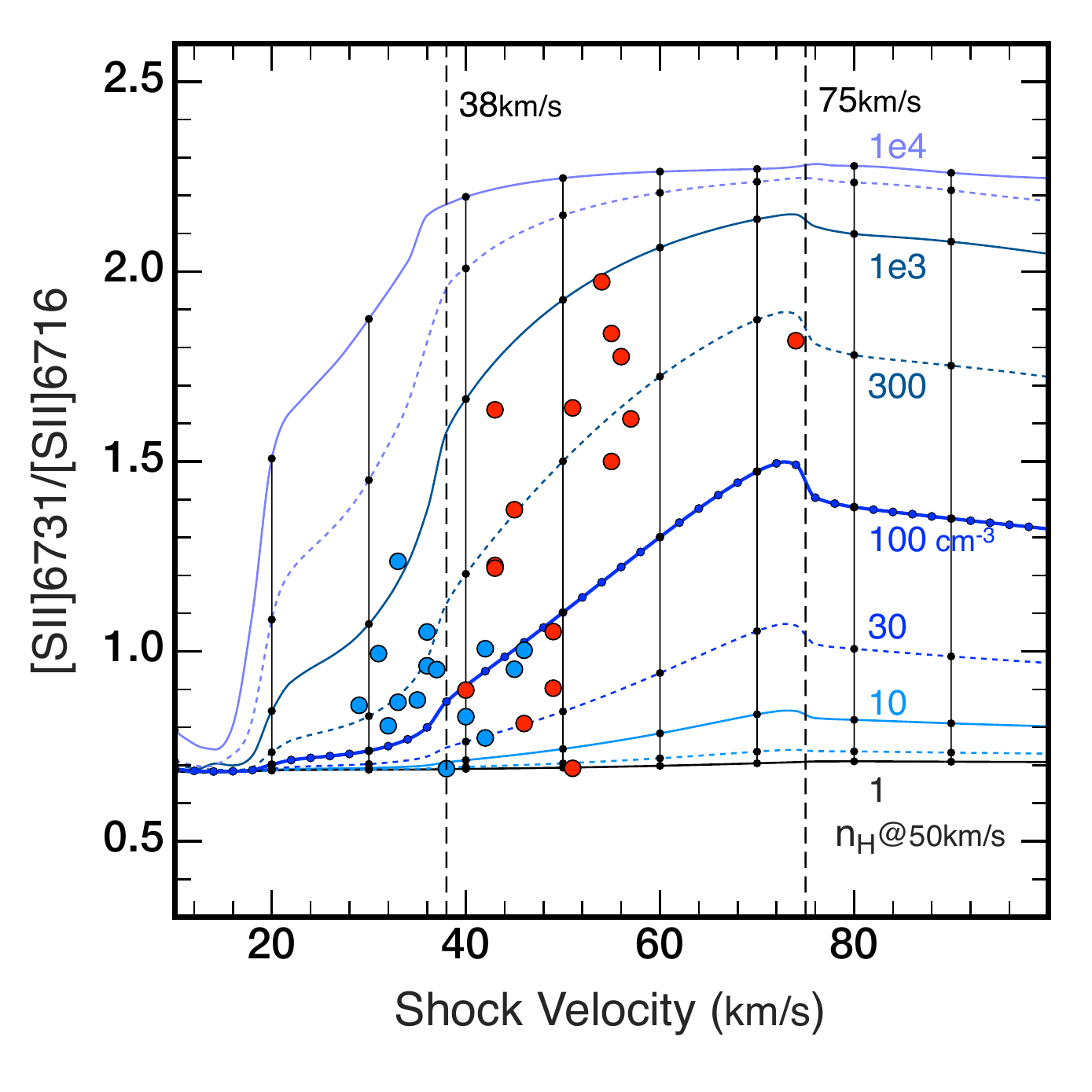}
  \caption{The [S\,II]$\lambda\lambda6717/6731$ line ratio provides a pressure diagnostic, provided that the shock velocity is known. Here we have used Figure \ref{fig:vel_diagnostic} to estimate the shock velocity for HH objects compiled by \citet{Raga96}, supplemented by the subsequent observations of \citet{Heathcote98}. The meaning of the symbols is the same as in Figure \ref{fig:diagnostics}.} \label{fig:press_diagnostic}
 \end{figure}

\begin{table}
\caption{Estimated global shock parameters of the HH objects using the diagnostics in this paper along with the spectrophotometric compilation of \citet{Raga96} and \citet{Heathcote98}. }
\label{Table:HHparms}
\begin{tabular}{lccc}
 \hline
{HH Object} & {$\log(P/k)$}  &  {$V_s$}  &  {$\log (n_1)$ } \\
                     & cm$^{-3}$K & km/s  & cm$^{-3}$ \\
\hline
& & & \\
{\bf  {[O\,III]/H$\beta < 0.15$~:}} & & & \\
HH7 & $7.44\pm0.23$$^{1}$ & $29\pm3$ & $2.32\pm0.14$\\
HH10 & $7.53\pm0.33$ & $31\pm5$ & $2.35\pm0.15$\\
HH11 & $7.22\pm0.20$ & $33\pm4$ & $1.99\pm0.05$\\
HH43B & $7.79\pm0.17$ & $37\pm3$  & $2.46\pm0.04$\\
HH43C & $7.02\pm0.30$ & $36\pm3$  & $1.74\pm0.06$\\
HH47A & $6.63\pm0.34$ &  $36\pm3$ & $2.19\pm0.13$\\
HH47C &  $6.53\pm0.34$ &  $42\pm3$ &$1.18\pm0.02$ \\
HH111V & $7.07\pm0.33$ &  $43\pm4$ & $1.70\pm0.02$ \\
HH111D-J & $7.31\pm0.15$ &  $33\pm3$ & $2.08\pm0.05$\\
HH125F & $<5.52$ & $46\pm7$ & $<0.0 $\\
HH128 & $7.20\pm0.017$ &  $32\pm3$ & $2.00\pm0.02$\\
HH54B & $7.03\pm0.05$ & $42\pm4$ & $1.57\pm0.05$\\
HH111L & $6.90\pm0.24$ & $40\pm3$ & $1.54\pm0.02$\\
HH83BD & $6.98\pm0.41$ & $45\pm7$ & $1.48\pm0.05$\\
HH34MD & $5.23\pm0.10$ & $38\pm3$ & $0.00\pm0.03$\\
& & & \\
{\bf {[O\,III]/H$\beta > 0.15$~:}} & & & \\
HH123 & $7.93\pm0.12$ & $43\pm3$ & $2.47\pm0.03$ \\
HH124A-C & $6.97\pm0.28$ & $49\pm6$& $1.41\pm0.08$ \\
HH24A & $7.58\pm0.15$ & $45\pm4$ & $2.08\pm0.05$ \\
HH158 & $8.19\pm0.07$ & $55\pm3$ & $2.50\pm0.04$ \\
HH240 & $7.35\pm0.11$ & $43\pm3$ &$1.89\pm0.11$ \\
HH54C &$6.89\pm0.08$  & $40\pm4$ & $1.48\pm0.02$ \\
HH234 & $6.60\pm0.32$ &  $46\pm6$ & $1.10\pm0.10$ \\
HH56 & $7.45\pm0.05$ &  $43\pm3$ & $1.99\pm0.03$ \\
HH3 & $8.56\pm0.28 $ &  $54\pm4$ & $2.92\pm0.10$ \\
HH43A & $7.13\pm0.23$  &  $49\pm3$ & $1.57\pm0.05$ \\
HH34apex & $5.55\pm0.31$  &  $51\pm4$ & $0.00\pm0.02$ \\
HH1 & $7.96\pm0.10$ &  $51\pm3$ & $2.33\pm0.07$ \\
HH2H & $8.08\pm0.07$  &  $56\pm3$ & $2.38\pm0.02$ \\
HH32 & $7.92\pm0.14$  & $57\pm3$ & $2.22\pm0.06$ \\
HH81 & $8.43\pm0.22$ & $74\pm6$ & $2.50\pm0.04$\\
HH80 & $7.67\pm 0.05$ & $55\pm5$ & $2.00\pm0.02$ \\
\hline
\end{tabular}
\newline
$^{1}$ Errors given are the fitting errors only, and do not include the photometric errors, which are not known..\\
\end{table}

\section{Physical Scales}
The density of the material being shocked, and the geometry of the shock itself determine the physical extent of the shock and its precursor. Clearly, these parameters determine the applicability of our computed shock models to real HH objects. In the following subsections we provide these essential quantities.

\subsection{Cooling Lengths}
It is useful to investigate the cooling length directly, since this determines the physical extent of the shock structure on the sky. However, at a given shock velocity, provided that collisional de-excitation effects are negligible, the cooling timescale (and therefore the cooling length) of the post-shock plasma varies inversely as the density. Thus the product of the cooling length and the pre-shock density -- which we term the cooling column, since it has dimensions of cm$^{-2}$ -- provides a quantity which, at the low density limit, is independent of density. 

In Paper I, we instigated this quantity over the full range of shock, density, and magnetic field parameter. When hydrogen is fully pre-ionised, for shock velocities above $v_s \sim 110$\,km/s, the cooling column can be represented by a single function. However, in the case of the HH objects, the variation in both the degree of pre-ionisation and the pre-shock temperature (which affect both the Mach number of the shock and the electron density at a given shock velocity and pre-shock density) invalidate a general scaling of this nature. In Figure \ref{fig:coolength}, we show both the computed cooling length (D$_3$; the distance to a final electron temperature of 1000\,K) and the cooling column ($\lambda_3$; again to a final electron temperature of 1000\,K) as a function of both shock velocity and pre-shock density.

The cooling length over the greatest range of interest for the HH objects ($20 < v_s \sim 75$\,km/s, actually decreases with shock velocity, thanks to the increased fractional ionisation reached in the shock, resulting in more efficient cooling. A 20\,km/s shock with a pre-shock density of $n_{\mathrm H} = 1000$\,cm$^{-3}$  has a cooling length of $10^{15}$\,cm. Using an assumed distance to the Orion Nebula Cluster of 400\,pc \citep{Kim16}, this corresponds to an angle of 0.17 arc sec. on the sky, which can just be resolved with HST. Using more typical values for the HH objects given in Table \ref{Table:HHparms} and Fig \ref{fig:diagnostics} ($v_s \sim 50$\,km/s, $n_{\mathrm H} = 100$\,cm$^{-3}$), the corresponding cooling distance is $\sim 1.6\times10^{14}$\,cm which corresponds to an angle of only 0.03 arc sec. on the sky at the distance of Orion.

\begin{figure*}
\centering
  \includegraphics[scale=0.6]{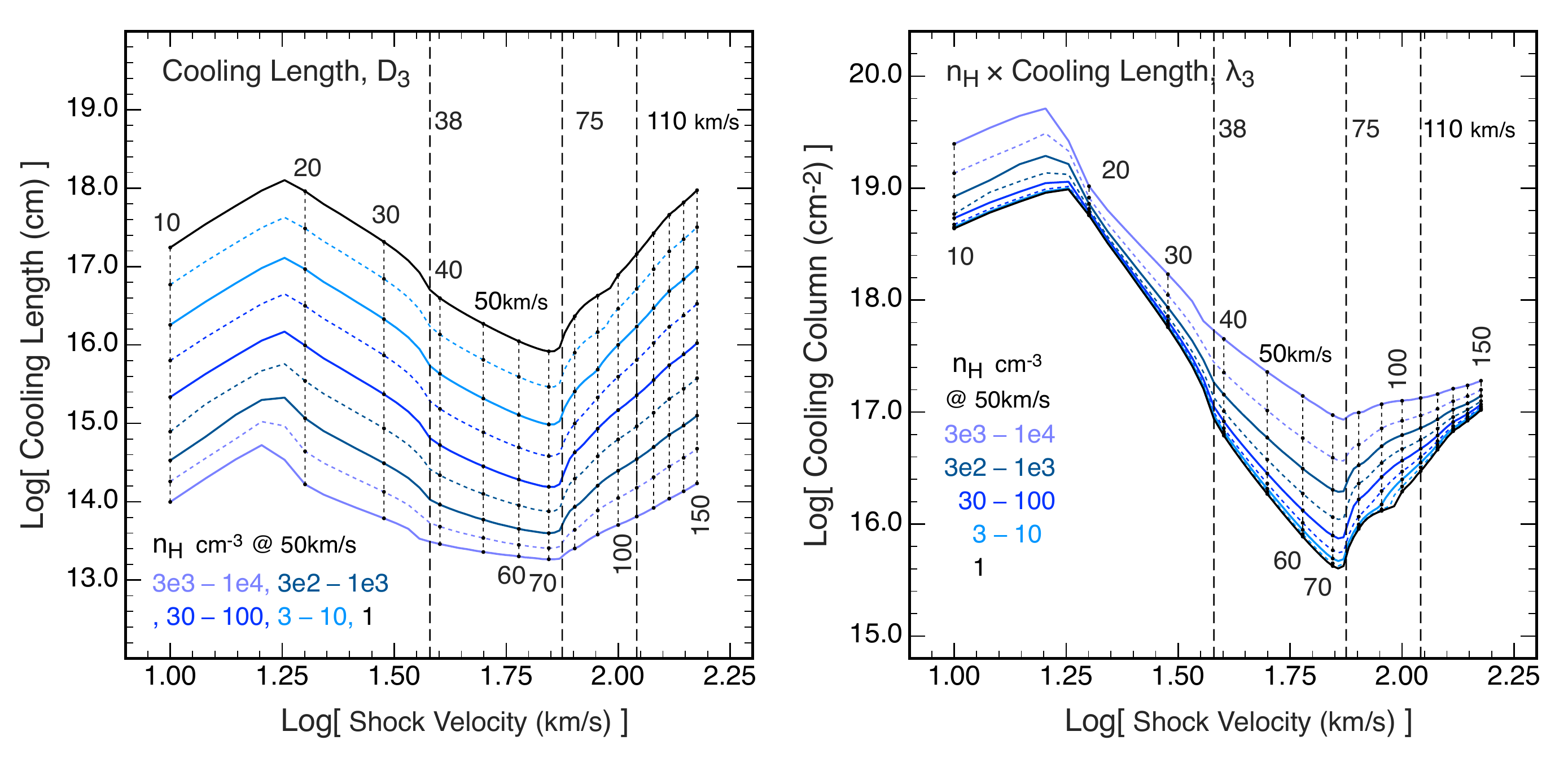}
   \caption{The cooling length (D$_3$),  the distance to a final electron temperature of 1000\,K) and the cooling column to a final electron temperature of 1000\,K ($\lambda_3$)  as a function of both shock velocity and pre-shock density.} \label{fig:coolength}
\end{figure*}

\subsection{Scale of the Precursor}
For shock velocities below 38\,km/s, the precursor is cool, and remains essential un-ionised. In this regime, what few ionising photons exist are absorbed in a column corresponding to the absorption edge opacity of atomic hydrogen ($\sim 10^{17}$\,cm$^{-2}$). For faster shocks, because the precursor fractional ionisation changes in a continuous manner ahead of the shock, the precursor region cannot be characterised by a single characteristic length. Instead, we choose to give the precursor lengths to initial H-ionisation fractions of 50\%, 10\%, 1\% and 0.1\%. These are shown as a function of shock velocity and pre-shock density on the left hand panel of Figure \ref{fig:preclength}. The left panel shows that  the full range of possible cooling lengths  may vary over 6 or more orders of magnitude from $\sim 10^{13}$  to over $10^{18}$ cm for shocks faster than $\sim 40$\,km/s. This rather complex panel is greatly simplified when we express the thickness of the precursor in terms of a Hydrogen column density, since this column density is independent of the physical density for any chosen value of the initial H-ionisation fraction (the right hand panel of Figure \ref{fig:preclength}).  Using typical values for the HH objects given in Table \ref{Table:HHparms} and Fig \ref{fig:diagnostics} ($v_s \sim 50$\,km/s, $n_{\mathrm H} = 100$\,cm$^{-3}$), the corresponding spatial scale of the precursor to an initial ionisation fraction of 0.1\%  is $\sim 3\times10^{15}$\,cm which corresponds to an angle of only $\sim 0.5$ arc sec. on the sky at the distance of Orion. However such a precursor would be very hard to detect. 

Note that the degree of ionisation inside the precursor increases relatively slowly as the shock is approached. This is a result of the time-dependent behaviour of the pre-ionisation, and is quite unlitke what would be expected for a precursor in photoionisation equilibrium. As a result, precursors, when visible, would be expected to appear rather ``fuzzy" and indistinct. Thus, the concept of a precusor length is rather poorly defined. Faster shocks will produce much stronger and more extensive precursors. In these the internal ionisation is much closer to the photoionisation equilibrium, and for shock velocities above $\sim 150$\,km/s the approximations made by \citet{Dopita96} are valid.

If we take the 10\% ionisation curves from Figure  \ref{fig:preclength} and compare these curves with the scale of the cooling region of the shock in 
Figure \ref{fig:coolength}, we see that the precursor length is always physically larger than the shock cooling length. In practice, this may mean that the precursor ISM runs out before a full precursor structure has developed, which in turn opens up the precursor to irradiation by external UV fields, briefly mentioned above in Section \ref{sec:SBprecursor}.

The precursors of plane-parallel shocks computed here give upper limits to the true strength of the precursor, since the upstream diffuse field is maximised in this geometry. In other geometries such as in bow shocks, the diffuse field is strong only at the leading edge, and geometrical dilution effects will reduce the degree of pre-ionisation in the Mach cone. This regime has been modelled in 3-D with ray-tracing methods used for the radiative transfer by \citet{Esquivel13}. We recommend the reader to read this paper and the earlier work by the UNAM team \citep{Raga94, Raga99} and by \citet{Raymond88}, which deal with the more complex shock geometries than the simple 1D plane parallel models presented here. A full treatment of time-dependent pre-ionisation in cases of complex geometry is beyond the scope of this paper.

\begin{figure*}
\centering
  \includegraphics[scale=0.6]{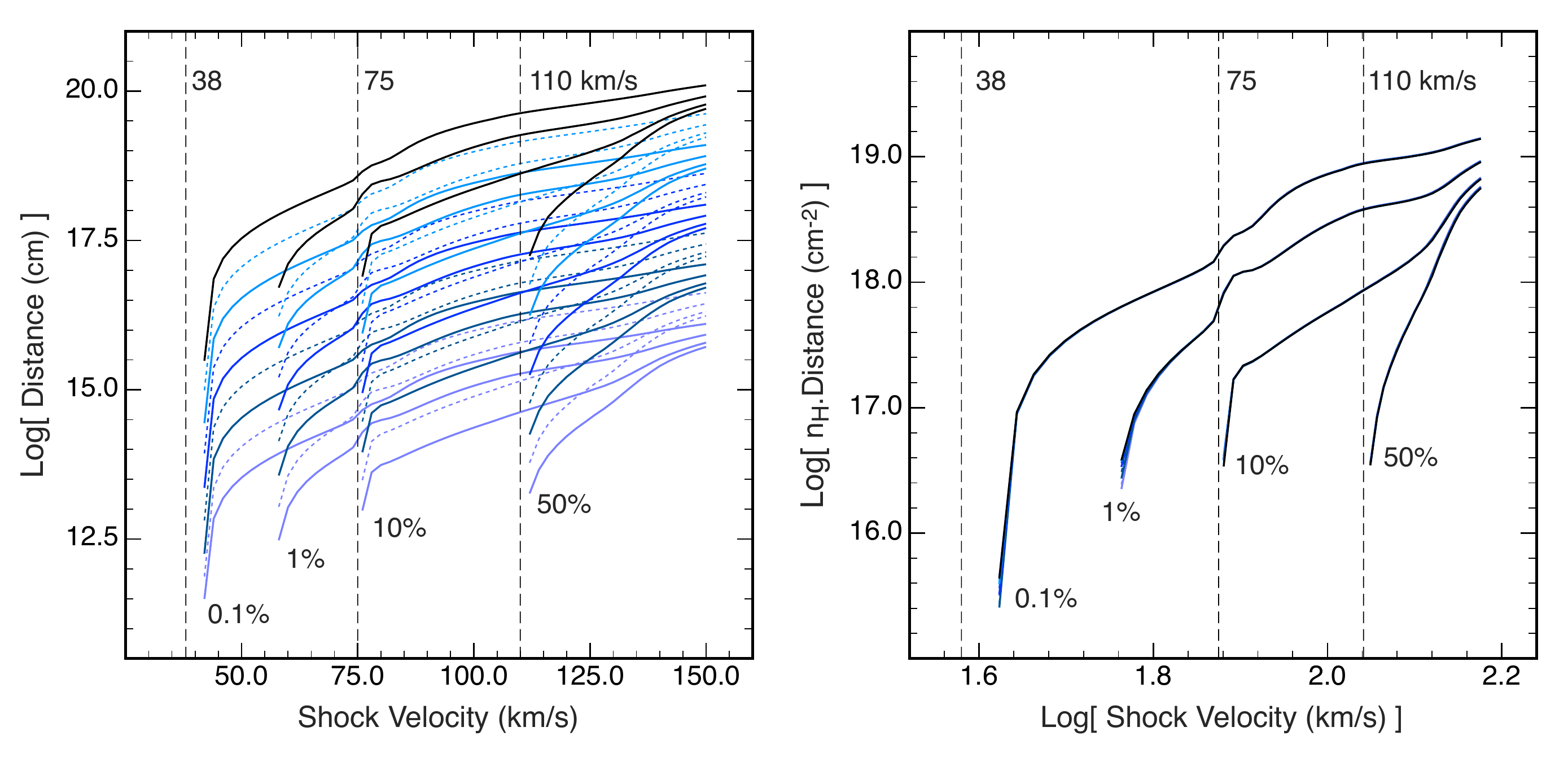}
   \caption{Left: The thickness of the precursor give at several limits to the pre-ionisation fraction as a function of both shock velocity and pre-shock density. The colour code for density is the same as in Figure \ref{fig:coolength}. Right: The H column density to these same fractional ionisations. These curves are effectively independent of the pre-shock density. } \label{fig:preclength}
\end{figure*}

\section{Application to HH34}
The HH34 jet and its associated multiple bow shocks presents a particularly fine example with which to test our shock models. The inner jet, which emerges from the star HH34 IRS, has a series of multiple shocked knots along its length which all have a remarkably low excitation \citep{Morse93} and very large proper motions \citep{Reipurth02}. The shocked knots are therefore internal to the jet, probably driven by velocity of mass ejection fluctuations in the outflow. In the IR, the jet displays very weak H$_2$ emission, but is very strong in [Fe\,II] \citep{Stapelfeldt91}. The jet is likewise characterised by very strong [Fe\,II], [Ni\,II] and [Ca\,II] emissions in the optical, suggesting that it is both atomic and dust-free in nature \citep{Morse93,Reipurth01}. The inner jet has already undergone detailed analysis in the paper by \citet{Hartigan94}, who demonstrated that the characteristic shock velocity is $\sim 28$ km/s, and that the ionisation fraction remains very low throughout the radiative shock structure. Based on an extensive grid of shock models, they derive both the mass loss rate and momentum flux in the jet.

The detailed structure in the HH34 jet has been revealed in the spectacular integral field study of \citet{Beck07}. They find the electron density as traced by the [S\,II]$\lambda\lambda 6731/6717$ ratio peaks on the leading edge of each of the emission knots in the collimated jet, consistent with the compression and increase in ionisation caused by a shock. The density in each knot decreases systematically with distance from the central star, consistent with the expansion of the jet observed by \citet{Reipurth01}. In addition, the radial velocity decreases with increasing distance from the central star, consistent with mass entrainment and the progressive dissipation of the mechanical energy flux of the jet through the action of internal working surface shocks along its length.  A similar behaviour as been found in the HH111 jet by \citet{Cerqueira15}. Their high spatial resolution \emph{Gemini MOS} IFU spectroscopy shows that the HH 111 jet has a fast, axial region with lower velocity shocks surrounded by a lower velocity sheath with higher velocity shocks.

The HH34 jet is merely the core of a much larger complex giant HH flow extending over 3pc  \citep{Devine97}. Some distance from the inner jet, the flow passes through a Mach disk structure, HH34, and this is enveloped in a higher-excitation bow shock, as seen clearly in the nice (colour) Hubble Space Telescope image of \citet{Reipurth02}. The counter-jet produces a similar structure HH34 N, which is followed further out by HH126, HH85 and HH33/40. On the jet side the corresponding regions are HH173, HH86, and HH87/88.

Optical spectrophotometry has been obtained by \citet{Morse93} for both the Mach disk (HH34 MD), and for the apex of the bow shock (HH34 apex) in HH34. However, a complete spectral separation between these two regions cannot be achieved as they are to a large degree co-spatial. Our shock modelling for these two regions will therefore use a linear sum of a lower velocity (Mach disk) shock, and a higher velocity (bow) shock emission.

\subsection{Modelling the HH34 jet}
In order to model the HH34 jet region, we note that the extremely high ratios of [O\,I] $\lambda\lambda 6300, 6363/H\alpha$ and [N\,I] $\lambda\lambda 5198, 5200/H\alpha$ imply a very low shock velocity in the range $22-30$\,km/s, as previously estimated by  \citet{Hartigan94}. As noted above, in this velocity range the collisional excitation of the Balmer lines is a major contributor the the Balmer line flux, and the Balmer decrement is, by consequence, much steeper than that predicted in Case B. We therefore first estimate the shock velocity from the excitation (c.f. Figure \ref{fig:diagnostics}), obtain the theoretical Balmer decrement appropriate to this velocity, and use this to determine the correct reddening correction for the \citet{Morse93} data with the \citet{Blagrave07} Orion region modified extinction law.  From this, we determined a logarithmic reddening $c_{\mathrm H\beta} = 0.2$ for the HH34 jet region. The de-reddened fluxes so obtained are given in Table \ref{table:HH34}. This same reddening correction was also applied to the two other regions of HH34, and the derived de-reddened fluxes are also given in the Table.

With the assumed abundance set, an extensive sub-grid of 100 models was run over a restricted range of shock velocity ($22.5-27.5$\,km/s,and shock ram-pressure (corresponding to $n_{\mathrm H} = 4800-8000$\,cm$^{-3}$ for a shock velocity of $25$\,km/s. These covered the vicinity of the best-fit solution. 

In order to measure the goodness of fit of any particular model, we measure the degree to which it reproduces the density-sensitive [S\,II] $\lambda\lambda 6731/6717$ ratio, and we also seek to minimise the L1-norm for the fit. That is to say that we measure the modulus of the mean logarithmic difference in flux (relative to H$\beta$) between the model and the observations \emph{viz.};
\begin{equation}
{\rm L1} =\frac{1}{m}{\displaystyle\sum_{n=1}^{m}} \left | \log \left[ \frac{F_n({\rm model})} {F_n({\rm obs.)}} \right]  \right |. \label{L1}
\end{equation}
This procedure weights fainter lines equally with stronger lines, and is therefore more sensitive to the values of the input parameters. We simultaneously fit at least 10 emission lines which are most sensitive to the shock velocity and ram pressure; effectively [O\,I], [O\,II] and if present, [O\,III] as well as the [N\,I], [N\,II], and [S\,II] lines. In addition, we fit the Balmer decrement H$\alpha$, H$\beta$ and H$\gamma$ and estimate the heavy element abundances using the [Fe\,II] 5158\AA\, [Ni\,II] 7378\AA\, Mg\,I] 4561\AA, Ca\,II 3933\AA\ and [Ca\,II] 7291\AA\ lines.

The fit so obtained is shown in Figure \ref{fig:HH34jet}. The shock velocity is very tightly constrained by the line flux ratios, $v_s = 25.0 \pm 1.2$\,km/s, but the pre-shock density has a larger error which is set by the photometric error in the observations. We estimate the pre-shock hydrogen density to be $n_{\mathrm H} = 6800^{+3500}_{-2000}$\,cm$^{-3}$.

In order to fit the lines of refractory elements, it is necessary to enhance the chemical abundances above the local concordance values. Indeed, all these elements, Mg, Ca, Fe and Ni must be enriched by approximately the same amount, +0.22\,dex, to obtain a satisfactory fit to the observations. The complete emission line predictions for our best-fit model ($v_s = 25$\,km/s, $n_{\mathrm H} = 6800$\,cm$^{-3}$) is given in Table \ref{table:HH34}.
\begin{figure}
\centering
  \includegraphics[scale=0.5]{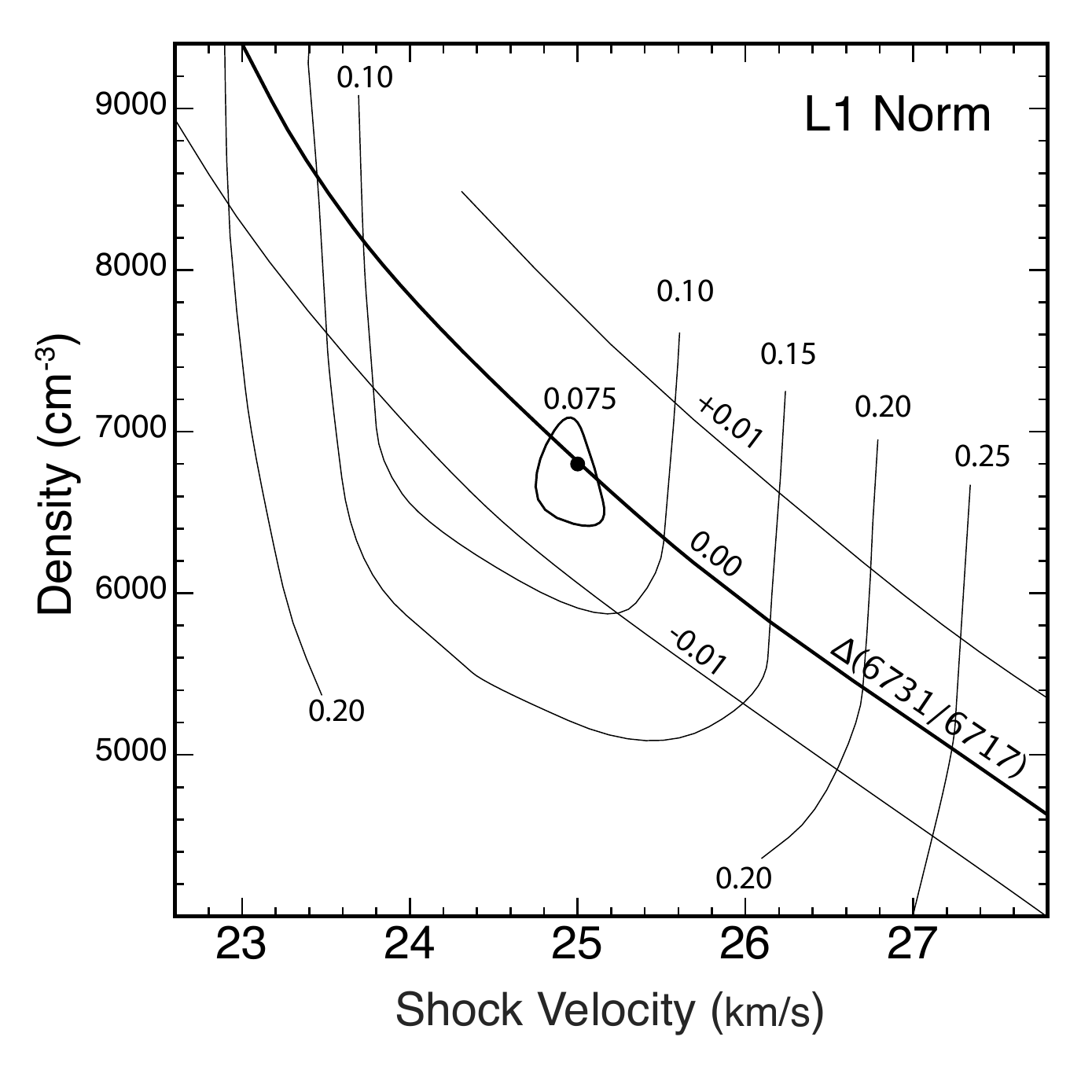}
   \caption{The goodness of fit contours for the HH34 jet region, derived from 100 shock models, plotted against the shock velocity and the pre-shock hydrogen density, $n_{\mathrm H}$. The L1-norm defines the shock velocity, $v_s = 25.0 \pm 1.2$\,km/s, while the  density-sensitive [S\,II] $\lambda\lambda 6731/6717$ ratio defines the pre-shock density, but with much larger error, set by the photometric error in the observations. We estimate $n_{\mathrm H} = 6800^{+3500}_{-2000}$\,cm$^{-3}$.} \label{fig:HH34jet}
\end{figure}

\subsection{HH34 Mach disk and Bowshock Regions}
As noted above, the Mach disk and the bowshock regions cannot be entirely separated in a spatial sense from one another. We have therefore adopted a model in which the observed spectrum results from a linear sum of a (high velocity) bow shock and a (lower velocity) cloud shock spectrum. To avoid introducing new variables, we adopt the chemical abundances determined for the HH43 jet. Since the ram pressure of the shock in the Mach disk drives the bowshock, the ram pressures in the two shocks are taken to be the same. The density in the Mach disk can be estimated from the density derived above for the HH34 jet, and from the ratio of the working surfaces of the Mach disk and the HH34jet shock. This ratio is derived from the HST images of \citet{Reipurth01} which show the jet to b 0.6\,arcsec. in diameter (at bright knots I and J), while the [S\,II] - bright Mach disk area covers 10\,arcsec. in diameter. The implied density of the Mach disk pre-shock region is $22^{+11}_{-7}$\,cm$^{-3}$. 

With this density, and matching the ram pressure in the Mach disk and bow shock, we varied the shock velocities, seeking to minimise the L1 norm in both regions. The bow shock has strong [O\,III], [O\,II] and [N\,II] emission relative to H$\beta$, but has weak [O\,I] and [N\,I], while most of the [O\,I] and [N\,I] emission arises in the Mach disk shock, so the contributions of the two can be readily separated. We find, for the Mach disk,  $v_s = 34.0 \pm 2$\,km/s and  $n_{\mathrm H} = 22^{+11}_{-7}$\,cm$^{-3}$ while for the bow shock, $v_s = 80 \pm 10$\,km/s and  $n_{\mathrm H} = 4^{+2}_{-1}$\,cm$^{-3}$. In the observations of HH34 MD, the Mach disk shock contributes around 80\% of the total H$\beta$ emission, while the HH34 apex region has only a $\sim30$\% contribution. Again, as for the HH34 jet region, the refractory elements, Mg, Ca, Fe and Ni are enriched by 0.22\,dex. The emission line fits are listed in Table \ref{table:HH34}.

\begin{table*}
 \centering
\caption{Best fit shock models for the three regions of HH34}\label{table:HH34}
\label{Table:HH34}
\begin{tabular}{lcccccccc}
 \hline
 \hline
Line ID and $\lambda$ (\AA):  &    HH34 jet \footnotemark[1] &    Model\footnotemark[2] \footnotemark[3]  &      &    HH34 MD \footnotemark[1]  &    Model\footnotemark[2] \footnotemark[4]   &      &    HH34 apex \footnotemark[1]   &    Model\footnotemark[2] \footnotemark[5] \\
\hline
{[O II] 3726,9}  &   {....} &    38.4  &      &    197.6  &    209.4  &      &    385.0  &    669.8\\
{[Ne III] 3868 }  &  {....} &    0.0  &      &   {....} &    1.0  &      &    14.6  &    3.4\\
{He I 3889, H$\zeta$ 3889}  &   {....} &      &      &      &    5.1  &      &    9.7  &    10.4\\
{Ca II 3933}  &    37.9  &    157.0  &      &    29.5  &    11.0  &      &    19.3  &    8.3\\
{[Ne III] 3967, Ca II 3968, H$\epsilon$ 3970} &   37.8  &    86.7  &      &    25.2  &    11.2  &      &    26.4  &    12.7\\
{[S II] 4069,76}  &    106.2  &    96.3  &      &    21.9  &    11.8  &      &    15.3  &    8.2\\
H$\delta$  4101&    14.5  &    12.0  &      &    18.6  &    11.3  &      &    24.6  &    16.0\\
{[Fe II]} 4244  &    39.0  &    46.9  &      &    4.1  &    5.8  &      &    4.5  &    2.9\\
{[Fe II]} 4277, 4287\footnotemark[6]   &    93.1  &    18.9  &      &    8.2  &    2.4  &      &    5.6  &    1.6\\
{H$\gamma$ 4340, [ Fe II] 4347}  &    45.8  &    49.1  &      &    30.5  &    39.9  &      &    44.6  &    41.8\\
{[O III] 4363,[Fe II] 4358}  &    46.6  &    17.4  &      &    8.9  &    2.4  &      &    7.8  &    4.2\\
{[Fe II] 4416}  &    37.3  &    45.8  &      &    7.7  &    5.7  &      &    4.4  &    2.9\\
{Mg I 4561,71}  &    145.2  &    135.0  &      &    23.3  &    16.2  &      &    10.6  &    7.4\\
{[Fe II] 4815}  &    30.3  &    36.6  &      &    5.0  &    4.4  &      &    2.0  &    2.2\\
{H$\beta$}  &    100.0  &    100.0  &      &    100.0  &    100.0  &      &    100.0  &    100.0\\
{[O III] 5007}  &   {....} &    0.0  &      &    13.7  &    13.7  &      &    52.3  &    47.1\\
{[Fe II] 5158 } & 152.1  &    167.2  &      &    18.1  &    20.7  &      &    6.7  &    10.4\\
{[N I] 5198,5200}  &    358.0  &    489.0  &      &    52.9  &    67.8  &      &    16.0  &    27.7\\
{[Fe II] 5262,[Fe III ]5270}  &    72.9  &    94.4  &      &    7.5  &    14.7  &      &    4.7  &    16.3\\
{[O I] 6300,64}  &    1466.5  &    1543.0  &      &    158.0  &    192.3  &      &    70.7  &    75.4\\
{H$\alpha$ 6563}  &    496.7  &    498.0  &      &    389.0  &    435.7  &      &    354.0  &    371.3\\
{[N II] 6583}  &    58.5  &    14.0  &      &    55.3  &    20.2  &      &    47.1  &    57.0\\
{[S II] 6717}  &    1649.9  &    1070.0  &      &    216.0  &    166.4  &      &    96.3  &    91.6\\
{[S II] 6731}  &    1441.6  &    932.0  &      &    149.1  &    115.4  &      &    66.5  &    65.1\\
{[Fe II] 7155}  &    200.5  &   {....} &      &    10.0  &   {....} &      &    3.9  &    {....}\\
{[Ca II] 7291 } & 478.2  &    342.0  &      &    65.5  &    30.1  &      &    22.1  &    20.2\\
{[Ca II] 7324, [O II] 7325}  &    179.1  &    248.0  &      &    40.2  &    30.6  &      &    27.3  &    49.3\\
{[Ni II] 7378}  &    240.2  &    169.1  &      &    10.6  &    6.1  &      &    4.5  &    5.0\\
\hline
\end{tabular}
\newline
$^{1}$ Reddening correction from \citet{Blagrave07} Orion region modified extinction law with logarithmic reddening $c_{\mathrm H\beta} = 0.2$.\\
$^{2}$ Local galactic concordance abundance with Mg, Si, Ca, Fe and Ni enhanced by 0.22\,dex.\\
$^{3}$ Shock velocity $v_s = 25$\,km/s, preshock density $n_{\mathrm H} =6800$\,cm$^{-3}$ \\
$^{4}$ 80\% Mach disk: $v_s = 34$\,km/s, $n_{\mathrm H} =21.6$\,cm$^{-3}$; 20\% \newline Bow Shock: $v_s = 80$\,km/s, $n_{\mathrm H} =3.9$\,cm$^{-3}$.\\
$^{5}$ 32\% Mach disk: $v_s = 34$\,km/s, $n_{\mathrm H} =21.6$\,cm$^{-3}$; 68\% \newline Bow Shock: $v_s = 80$\,km/s, $n_{\mathrm H} =3.9$\,cm$^{-3}$.\\
$^{5}$Flux of [Fe II] $\lambda 4287$ not predicted in MAPPINGS.\\
  \end{table*}

\subsection{The physical parameters of HH34}
With the derived shock parameters, we can estimate a number of interesting parameters for the HH34 jet and for its Mach disk. Firstly, assuming that the jet is two-sided, we can estimate the mass-loss rate in the jet; $\dot M = (2\pi/4)\mu m_{H} D^2 v$ or
\begin{equation}
\dot M = 3 \times 10^{-8} \left[\frac {v}{250km\,s^{-1}}\right ]\left[\frac{n_H}{10^4cm^{-3}}\right]\left[\frac{D}{100AU}\right]^2 {\mathrm M_{\odot}yr^{-1}}
\end{equation}
where $n_{H}$ is the hydrogen particle density in the jet, $v$ its velocity, and $D$ its diameter. From  \citet{Reipurth01} we have $D \sim 250$\,AU in the vicinity of the bright knots G,H and I. These authors estimate the space velocity of these knots to be $\sim 185$\,km/s. To this we must add the derived shock velocity to obtain the bulk velocity $v = 210$\,km/s. With our pre-shock density, $n_{\mathrm H} = 6800$\,cm$^{-3}$ , we have $\dot M = 1.1 \times 10^{-7}{\mathrm M_{\odot}}$yr$^{-1}$. This compares well with the \citet{Hartigan94} value of $1.5 \times 10^{-7}{\mathrm M_{\odot}}$yr$^{-1}$, but is somewhat smaller than the value of $3.7 \times 10^{-7}{\mathrm M_{\odot}}$yr$^{-1}$ derived by \citet{Bacciotti99}. The momentum flux (or thrust) of $\dot M v = 3.7 \times 10^{-5}{\mathrm M_{\odot}}$\,km/s\,yr$^{-1}$, likewise similar to the $\dot M v = 4 \times 10^{-5}{\mathrm M_{\odot}}$\,km/s\,yr$^{-1}$ derived by \citet{Hartigan94}, but again somewhat smaller than the \citet{Bacciotti99} estimate;  $\dot M v = 8.1 \times 10^{-5}{\mathrm M_{\odot}}$\,km/s\,yr$^{-1}$. These differences are driven mainly by different author's estimates of the pre-shock density. For example, we derived $n_{\mathrm H} = 6800$\,cm$^{-3}$, whereas \citet{Bacciotti99} obtained $n_{\mathrm H} = 2\times10^4$\,cm$^{-3}$ from their analysis. These differences are in turn driven by the estimated degree of pre-ionisation, and by the detailed structure of the radiative shock, both of which determine the measured electron density, which is the physical quantity that is used to estimate the pre-shock density.

The sideways expansion of the jet (assumed to be into a vacuum) is $v_{exp} = c_{1}\left[2/(\gamma -1)\right]^{1/2}$, where $c_{1}$ is the sound speed in the pre-shock gas, and $\gamma = 5/3$ is its adiabatic index. From the self-consistent pre-ionisation state computed in our model, we find a pre-shock temperature of $T=280$K, corresponding to a sound speed of 0.5\,km/s. Thus over the $\sim 200$yr age of the jet (out to bright knots G,H and I) the jet would have expanded to a diameter of 140AU assuming thermal expansion acting alone. This is somewhat smaller than the observed diameter, 250AU. However, there is certainly some contribution to the transverse expansion due to post-shock turbulence in the jet. We must conclude that the jet was originally extremely well-collimated.

The enhanced abundances of the refractory elements tell us something about the launch region of the jet. It is clear that, not only must this lie within the grain sublimation radius, in order to return Ca, Fe and Mg to the gas phase, but also that the material being ejected has been chemically enhanced by having a greater fraction of its material derived from a dusty or rocky origin. Adopting a stellar luminosity ${\mathrm L}_* \sim 45 {\mathrm L}_{\odot}$ \citep{Cohen87}, and assuming a grain sublimation temperature $T_{sub}\sim 1000$K, of we can estimate the  grain sublimation radius; $r_{sub} = \left[ {\mathrm L}_* / 4\pi \sigma T_{sub}^4 \right ]$. Thus we can conclude that the jet is launched within a radius of  1.0\,AU of the star, which is well within the zone of the rocky planets. If the (heavy element enhanced) material seen in the jet is of the same composition as the material currently accreting onto the star, then we can expect that the final photospheric abundances may be enhanced above the interstellar abundances from which the star originally formed. 

The fact that the bow shock in HH34 has a shock velocity $\sim 80$\,km/s, but a has proper motion and radial velocity consistent with a space velocity of  $260-320$\,km/s \citep{Devine97, Reipurth02} reveals that the substrate medium involved in the bow shock is already outflowing at a space velocity of $180-240$\,km/s, and is therefore is part of the material ejected by HH34 IRS. This is consistent with the measured chemical abundances which are the same as in HH34 jet. We would expect that, as entrainment of the ambient ISM into this outflowing material becomes increasingly significant, the velocities of outflow in each of the HH complexes in the giant flow (on the north side HH34 N, HH126, HH85 and HH33/40, and on the south side, HH34, HH173, HH86, and HH87/88) will systematically decrease. This is consistent with the observed spatial velocities of these complexes which have been measured by \citet{Devine97}.\newpage

\section{Conclusions}
An understanding of the pre-ionisation conditions of low-- and intermediate--velocity shocks is necessary for the proper interpretation spectra of  HH Objects. In this paper, we have presented self-consistent models which permit us to estimate shock velocities, pre-shock densities and shock ram pressures using existing spectrophotometry. In particular we have applied these to a detailed modelling of the HH34 jet and Mach disk system, from which we have been able to derive physical parameters of the jet and outflow, and estimate the chemical abundances in this material. This study serves to show that shock modelling is vital if we are to understand the nature of outflows in pre-main sequence stellar systems, and the parameters derived for these outflows in turn informs studies aimed at understanding accretion and planetary system formation around these stars. 

Currently, a major restriction to the application of this type of modelling is the fact that much of the spectrophotometry is old, does not cover the required wavelength range, is often taken with single-slit spectrographs at too low a spectral resolution, and has been de-reddened using inappropriate Balmer decrements. Therefore, more integral field spectroscopy such as has been obtained by  \citet{Beck07} and \citet{Lopez10} is sorely needed. With such data, the 2D structure of the jet can be probed, and the different shock components (working surface shocks, Mach disk, bow shocks and cocoon shocks) can be separately analysed. For this purpose, the Wide Field Spectrograph (WiFeS) \citep{Dopita07} is ideally suited, since it can cover 3500--9500\AA\ at a resolution of $R=3000$ over a $38\times25$\,arc sec. field. Studies of HH jets with this instrument are proceeding, and the analysis of this data will be the subject of future papers.

\section*{Acknowledgments}
We thank the referee for an expert review of our paper, which has helped us improve it by both adding essential material on the spatial scales of these shocks, and in clarifying the physics.. The authors acknowledges the support of the Australian Research Council (ARC) through Discovery project DP16010363. \newpage

%\bibliographystyle{mn2e_new}
%\bibliography{refs}

%%\bibliographystyle{aasjournal}
%%\bibliography{refs}
%%\end{document}

%%%%%%%%%%%%%%%%%%%%%%%%%%%%%%%%%%%%%%%%%%%%%%%%%

% Don't change these lines
\end{document}